\documentclass[osajnl,twocolumn,showpacs,groupaddress,10pt]{revtex4-1} 
\usepackage{amsmath,amssymb,graphicx}
\usepackage{epstopdf}
\usepackage[utf8]{inputenc}
\usepackage{float}

\begin{document}

\title{Optical properties of multilayer optics\\including negative index materials}

\author{Michel Lequime}\email{Corresponding author: michel.lequime@fresnel.fr}
\author{Boris Gralak}
\author{Sébastien Guenneau}
\author{Myriam Zerrad}
\author{Claude Amra}
\affiliation{Aix Marseille Université, CNRS, Centrale Marseille, Institut Fresnel, UMR 7249, 13013 Marseille, France}

\begin{abstract}
Negative indices are revisited through the thin-film admittance formalism. Effective indices and phase delay associated with wave propagation through negative index layers are carefully defined and computational rules easily implementable in standard thin-film software are derived from this approach. This admittance formalism is then used to recover the main features of the perfect lens and to highlight the benefit of such negative index materials to improve the performances of quarter-wavelength Bragg mirrors and Fabry-Perot band-pass filters.
\end{abstract}

\maketitle

\section{Introduction}

In the field of electromagnetic optics, the past fifteen years have seen the emergence of new and astounding concepts directly connected with the potential availability of artificial materials, such as nano-engineered materials \cite{Shalaev_2007,Jiang_2013} or photonic crystals \cite{Gralak_2000,Notomi_2000}. These metamaterials are rationally designed to exhibit unprecedented optical properties like negative \cite{Veselago_1968,Shelby_2001} or zero \cite{Kocaman_2011} index of refraction, which create ways to achieve innovative functionalities unattainable with natural materials, as the perfect flat lens \cite{Pendry_2000} or the invisibility cloak \cite{Pendry_2006,Zolla_2007,Schurig_2006,Ergin_2010}.

Beside the first preliminary experimental demonstrations of these new and exciting concepts, the thin-film community was used to play with tailored "arbitrary indices in the complex plane" and to build complex optical functions based on their use: the implementation of the admittance formalism \cite{Macleod_2010} was the main mathematical tool to allow for such tailoring. However, despite this favorable context, negative indices have not really been disseminated within this community.

The main objective of this work is to help rectify this situation, by adapting the concept of negative index to the admittance theory, and then by using this extended formalism to analyze the optical properties of some standard multilayer stacks, like the quarter-wavelength Bragg mirror, the Fabry-Perot bandpass filter, or the antireflection coating, where one or more layers of these stacks involve Negative Index Materials (NIMs).

\section{Negative index of refraction}
\label{sec:NegativeIndexofRefraction}

In general, the refractive index of a medium at a frequency $\omega$ is given from Maxwell's equations by the following relation
\begin{equation}
n^2(\omega)=\epsilon_r(\omega)\mu_r(\omega)=\frac{\tilde{\epsilon}(\omega)\tilde{\mu}(\omega)}{\epsilon_0\mu_0}
\label{eq:RefractiveIndex}
\end{equation}
where $\tilde{\epsilon}(\omega)$ and $\tilde{\mu}(\omega)$ are the Fourier transforms of the temporal electric permittivity $\epsilon(t)$ and magnetic permeability $\mu(t)$ of the medium, and $\epsilon_0$, $\mu_0$ are the electromagnetic properties of vacuum. Equation (\ref{eq:RefractiveIndex}) is valid for linear and isotropic media, and the Fourier transform involves an exponential in the form $\exp(i\omega t)$.

These frequency dependent quantities are complex, hence the refractive index $n$ is not a well-defined single-valued function. Therefore, the computation of the square root of $\epsilon_r(\omega)\mu_r(\omega)$ requires the choice of a branch (also called a Riemann sheet \cite{Ramakrishna_2005}), that is a portion of its range over which this function is single-valued. For the square root function, the two possible principal values differ only by their signs. Although the choice of this sign for the refractive index of left-handed metamaterials has been a subject of controversy \cite{Chen_2005,*Mackay_2006,*Ramakrishna_2007, *Chen_2007,Stockman_2007,*Mackay_2007,Govyadinov_2007}, all the proposed approaches provide identical conclusions for media that are both negatively refracting and nonamplifying. As shown in next paragraphs, the choice of the negative sign leads to conditions (on the permittivity and permeability) derived from Maxwell's equation's and causality. 
A first set of relations is 
\begin{align}
&\epsilon_r'' \ge 0\text{ ; }\mu_r'' \ge 0 \label{eq:DissipativeRequirement}\\
&\epsilon_r' \mu_r'' + \epsilon_r'' \mu_r' \le 0 \label{eq:AmplitudeDecay}
\end{align}
where prime and double prime denote the real and imaginary parts of each complex quantity, respectively. The relation (\ref{eq:DissipativeRequirement}) is simply the dissipative requirement corresponding to an absorbing or transparent medium. As to the relation (\ref{eq:AmplitudeDecay}), it plays a vital role since, as shown hereafter, it ensures that a plane wave decays in amplitude as it propagates in a dissipative NIM medium. Finally, in the low-loss regime, the combination of these relations yields the well-known conditions
\begin{equation}
\epsilon_r' < 0\text{ ; }\mu_r' < 0
\label{eq:NIM}
\end{equation}

In order to understand the link between these last two relations and the wave amplitude decay in a NIM medium, we first compute the square roots of a complex number $z$ ($z=x+iy$) by using the following general formula \cite{Cooke_2008} 
\begin{equation}
\left[\sqrt{z}\right]_{\pm}=\pm\left[\sqrt{\frac{|z|+x}{2}}+i\thinspace\text{sgn}(y)\sqrt{\frac{|z|-x}{2}}\right]
\label{eq:ComplexSquareRoot}
\end{equation}
where $\text{sgn}(y)$ corresponds to the sign of the imaginary part $y$.
Then, by introducing the real and imaginary parts of each quantity in equation (\ref{eq:RefractiveIndex}), we infer
\begin{equation}
\left\{
\begin{aligned}
&x=\epsilon_r'\mu_r'-\epsilon_r''\mu_r''\\
&y=\epsilon_r'\mu_r''+\epsilon_r''\mu_r'
\label{eq:xy}
\end{aligned}
\right.
\end{equation}
and thus, for the refractive index $n$
\begingroup
\small
\begin{equation}
\left\{
\begin{aligned}
&n^{\prime}=\pm\sqrt{\frac{r+(\epsilon_r'\mu_r'-\epsilon_r''\mu_r'')}{2}}\\
&n^{\prime\prime}=\pm\text{sgn}(\epsilon_r'\mu_r''+\epsilon_r''\mu_r')\sqrt{\frac{r-(\epsilon_r'\mu_r'-\epsilon_r''\mu_r'')}{2}}
\end{aligned}
\right.
\label{eq:nprimendoubleprime}
\end{equation}
\endgroup
where $r=|z|$ is given by
\begin{equation}
r=\sqrt{(\epsilon_r'\mu_r'-\epsilon_r''\mu_r'')^2+(\epsilon_r'\mu_r''+\epsilon_r''\mu_r')^2}
\end{equation}
For a plane wave propagating in the positive $z$ direction in a NIM medium, the phase and amplitude changes are driven by an exponential in the form 
\begin{equation}
e^{ik_0nz}=e^{ik_0n^{\prime}z}\thinspace e^{-k_0n^{\prime\prime}z}
\end{equation}
where  $k_0$ is the modulus of the wave-vector in vacuum.

Thus, for a NIM medium ($n^{\prime}<0$), we must choose the minus sign in the general square root formula (\ref{eq:ComplexSquareRoot}), while the decay of the wave amplitude during the propagation imposes that the condition that $n^{\prime\prime}$ be positive or equal to zero, which leads to
\begin{equation}
\epsilon_r'\mu_r''+\epsilon_r''\mu_r'\leqslant 0
\notag
\end{equation}
Moreover, in the low-loss regime, we can write, as a first approximation
\begin{equation}
r\approx |\epsilon_r'\mu_r'-\epsilon_r''\mu_r''|=|x|
\end{equation}
If the product $\epsilon_r'\mu_r'$ was negative, $x$ would be negative too, following (\ref{eq:DissipativeRequirement}) and (\ref{eq:xy}), and $n^{\prime}$ would be equal to zero, in accordance with (\ref{eq:nprimendoubleprime}) taken at the same level of approximation. This means that the product $\epsilon_r'\mu_r'$ must be positive: The two real parts $\epsilon_r'$ and $\mu_r'$ are thus of the same sign, and from conditions (\ref{eq:DissipativeRequirement}) and (\ref{eq:AmplitudeDecay}), we can conclude that they are simultaneously negative, see (\ref{eq:NIM}), whereas the refractive index $n$ is given, in the same low-loss regime, by \cite{Ramakrishna_2005_2}
\begin{equation}
n_{\text{NIM}}\approx-\sqrt{\epsilon_r'\mu_r'}-i\frac{\epsilon_r'\mu_r''+\epsilon_r''\mu_r'}{2\sqrt{\epsilon_r'\mu_r'}}
\end{equation}

\section{Admittance formalism}

\subsection{General presentation}
Let us now consider a multilayer stack deposited at the surface of a plane substrate and illuminated by a monochromatic plane wave at an angle of incidence (AOI) $\theta_0$. At any position within the stack, the tangential component of the magnetic field $\vec{\mathcal{H}}_{\text{tan}}$ and the cross product between the normal to the substrate $\vec{z}$ and the tangential component of the electric field $\vec{\mathcal{E}}_{\text{tan}}$ are parallel \cite{Macleod_2010,Amra_1993}. The algebraic ratio of these quantities is called the complex admittance $Y$. Hence
\begin{equation}
\vec{\mathcal{H}}_{\text{tan}}=Y(\vec{z}\wedge\vec{\mathcal{E}}_{\text{tan}})
\end{equation}
If we are able to compute the complex admittance $Y_0$ of the entire stack, we can easily determine the amplitude reflection coefficient $r$ of the plane wave on the stack by using the formula
\begin{equation}
r=\frac{\tilde{n}_0-Y_0}{\tilde{n}_0+Y_0}
\end{equation}
where $\tilde{n}_0$ is the effective index of the incident medium, i.e. the proportionality factor between the same vectors as previously stated, but for the incoming plane wave.

The computation of the $Y_0$ factor is based on the application of a recursive formula linking the admittances at two consecutive boundaries \cite{Amra_1993}
\begin{equation}
Y_{j-1}=\frac{Y_j\cos\delta_j-i\tilde{n}_j\sin\delta_j}{\cos\delta_j-i(Y_j/\tilde{n}_j)\sin\delta_j}
\label{eq:RecurrenceRelation}
\end{equation}
where $j$ is the layer number, $\tilde{n}_j$ its effective index and $\delta_j$ the phase delay introduced by the crossing of this layer. The initialization of this recursive formula is made in the substrate where only the outgoing plane wave is present
\begin{equation}
Y_p=\tilde{n}_s
\label{eq:RecurrenceInitialization}
\end{equation}
where $p$ is the total number of layers in the stack and $\tilde{n}_s$ the effective index of the substrate.

Therefore, to define the optical properties of a multilayer stack including a negative index material, we first need to determine the \textbf{effective index} $\tilde{n}_{\text{NIM}}$ of this artificial material, and then the \textbf{phase delay} $\delta_{\text{NIM}}$ associated with the crossing of a layer thickness $t_{\text{NIM}}$.

\subsection{NIM effective indices}
\label{sec:NIMEffectiveIndices}
In general, the effective indices of a medium are given by \cite{Amra_1993}
\begin{equation}
\tilde{n}=\left\{
\begin{aligned}
&\alpha/\omega\mu_0\mu_r\quad\text{for TE polarization}\\
&\omega\epsilon_0\epsilon_r/\alpha\quad\enskip\text{for TM polarization}
\end{aligned}
\right.
\end{equation}
where $\alpha$ is defined by
\begin{equation}
\alpha^2=k^2-\sigma^2
\label{eq:alpha2}
\end{equation}

If we first consider the propagation of a plane wave within a transparent positive index material (PIM), we have 
\begin{equation}
\sigma=k\sin\theta
\end{equation}
and then
\begin{equation}
\alpha=k\cos\theta=\omega\sqrt{\epsilon_0\mu_0}\sqrt{\epsilon_r\mu_r}\cos\theta
\end{equation}
For a plane wave passing through a multilayer stack, $\sigma$ is an invariant while $\alpha$ depends on the layer medium. Therefore, the effective indices of a PIM medium are defined by
\begin{equation}
\left\{
\begin{aligned}
&\tilde{n}_{\text{PIM}}^{\text{TE}}=\frac{1}{\eta_0\mu_r}\thinspace n\cos\theta\\
&\tilde{n}_{\text{PIM}}^{\text{TM}}=\frac{1}{\eta_0\mu_r}\thinspace\frac{n}{\cos\theta}
\end{aligned}
\right.
\end{equation}
where $\eta_0=\sqrt{\frac{\mu_0}{\epsilon_0}}$ is the vacuum impedance.

For a negative index material illuminated by a plane wave at an angle of incidence $\theta_0$, relation (\ref{eq:alpha2}) becomes
\begin{equation}
\alpha^2=\omega^2\epsilon_0\mu_0(\epsilon_r\mu_r-n_0^2\sin^2\theta_0)
\end{equation}
where $n_0$ is the refractive index of the incident medium. To compute $\alpha$, we again need to obtain the square root of a complex number whose real and imaginary parts in this case are defined by
\begin{equation}
\left\{
\begin{aligned}
&\Re\left[\frac{\alpha^2}{\omega^2\epsilon_0\mu_0}\right]=(\epsilon_r'\mu_r'-n_0^2\sin^2\theta_0)-\epsilon_r''\mu_r''\\
&\Im\left[\frac{\alpha^2}{\omega^2\epsilon_0\mu_0}\right]=\epsilon_r'\mu_r''+\epsilon_r''\mu_r'
\label{eq:xyalpha}
\end{aligned}
\right.
\end{equation}
To apply here an approach strictly identical to the one detailed in Section \ref{sec:NegativeIndexofRefraction} for a NIM medium in the low-loss regime, we need to satisfy an additional condition
\begin{equation}
\epsilon_r'\mu_r'-n_0^2\sin^2\theta_0\gg\epsilon_r''\mu_r''
\label{eq:AdditionalCondition}
\end{equation}
which allows us to give an approximate expression for $\alpha$, i.e.
\begin{multline}
\alpha_{\text{NIM}}\approx-\omega\sqrt{\epsilon_0\mu_0}\thinspace\sqrt{\epsilon_r'\mu_r'-n_0^2\sin^2\theta_0}\\
-i\omega\sqrt{\epsilon_0\mu_0}\thinspace\frac{\epsilon_r'\mu_r''+\epsilon_r''\mu_r'}{2\sqrt{\epsilon_r'\mu_r'-n_0^2\sin^2\theta_0}}
\end{multline}
Moreover, in the transparency regime ($\epsilon_r''=\mu_r''=0$), relation (\ref{eq:AdditionalCondition}) can be rewritten as a \textit{propagating condition}
\begin{equation}
\epsilon_r'\mu_r'-n_0^2\sin^2\theta_0\geqslant 0
\end{equation}
and the effective indices of the NIM medium are thus given by
\begin{equation}
\left\{
\begin{aligned}
&\tilde{n}_{\text{NIM}}^{\text{TE}}=-\frac{1}{\eta_0\mu_r'}\sqrt{\epsilon_r'\mu_r'-n_0^2\sin^2\theta_0}\\
&\tilde{n}_{\text{NIM}}^{\text{TM}}=-\frac{\epsilon_r'}{\eta_0}\frac{1}{\sqrt{\epsilon_r'\mu_r'-n_0^2\sin^2\theta_0}}
\end{aligned}
\right.
\label{eq:NIMEffectiveIndices}
\end{equation}

The effective indices of a transparent NIM layer ($\epsilon_r'<0$, $\mu_r'<0$) are then \textbf{positive} in the \textit{propagating mode}.

Besides from that, by using the invariance of $\sigma$, we can replace in (\ref{eq:NIMEffectiveIndices}) $n_0^2\sin^2\theta_0$ by $\epsilon_r'\mu_r'\sin^2\theta$ to obtain
\begin{equation}
\left\{
\begin{aligned}
&\tilde{n}_{\text{NIM}}^{\text{TE}}=-\frac{1}{\eta_0\mu_r'}\sqrt{\epsilon_r'\mu_r'}\cos\theta=\frac{1}{\eta_0\mu_r}\thinspace n\cos\theta\\
&\tilde{n}_{\text{NIM}}^{\text{TM}}=-\frac{\epsilon_r'}{\eta_0}\frac{1}{\sqrt{\epsilon_r'\mu_r'}\cos\theta}=\frac{1}{\eta_0\mu_r}\thinspace\frac{n}{\cos\theta}
\end{aligned}
\right.
\end{equation}
The relations defining the effective indices of a NIM layer are thus seen to be identical to those obtained for the effective indices of a PIM layer.

\subsection{NIM phase delay}

Making use of the results from the previous section, we can easily compute a general expression for the phase delay $\delta_{\text{NIM}}$ associated with a NIM layer thickness $t_{\text{NIM}}$ in the \textit{propagating mode} of the transparency regime
\begin{multline}
\delta_{\text{NIM}}=\alpha_{\text{NIM}}\thinspace t_{\text{NIM}}\\=-\omega t_{\text{NIM}}\sqrt{\epsilon_0\mu_0}\thinspace\sqrt{\epsilon_r'\mu_r'-n_0^2\sin^2\theta_0}\\
=\frac{2\pi}{\lambda}(-t_{\text{NIM}})\thinspace\sqrt{\epsilon_r'\mu_r'-n_0^2\sin^2\theta_0}
\label{eq:DeltaNIM}
\end{multline}
Using the same approach as in Section \ref{sec:NIMEffectiveIndices} (replacement of $n_0^2\sin^2\theta_0$ by $\epsilon_r'\mu_r'\sin^2\theta$), we can rewrite relation (\ref{eq:DeltaNIM}) in the form
\begin{multline}
\delta_{\text{NIM}}=\frac{2\pi}{\lambda}(-t_{\text{NIM}})\thinspace\sqrt{\epsilon_r'\mu_r'}\cos\theta\\
=\frac{2\pi}{\lambda}(t_{\text{NIM}})\thinspace n\cos\theta
\end{multline}

\subsection{PIM Mirror Image}
\label{sec:PIMMirrorImage}

Let us consider a transparent NIM layer and a standard positive index material (PIM) whose electromagnetic properties (relative electric permittivity and relative magnetic permeability) are equal and opposite to the NIM ones
\begin{equation}
\epsilon_{r,\text{NIM}}^{\prime}=-\epsilon_{r,\text{PIM}}^{\prime}\text{ ; }\mu_{r,\text{NIM}}^{\prime}=-\mu_{r,\text{PIM}}^{\prime}
\end{equation}
In the propagating mode of the transparency regime, one can write
\begin{equation}
\left\{
\begin{aligned}
&\tilde{n}_{\text{NIM}}=\tilde{n}_{\text{PIM}}\\
&\delta_{\text{NIM}}=\alpha_{\text{NIM}}\thinspace t_{\text{NIM}}=-\alpha_{\text{PIM}}\thinspace t_{\text{NIM}}
\end{aligned}
\right.
\end{equation}
This last equation shows that we can simulate the properties of a NIM layer by replacing it with an equivalent PIM layer, provided that we use for this PIM layer a \textbf{virtual} thickness \textbf{opposite} to that of the NIM layer ($t_{\text{PIM}}=-t_{\text{NIM}}$). We now call this layer the correlated PIM \textit{Mirror Image} (PIMMI) of the NIM layer.

This property is essential because it allows us to use standard thin-film software to predict, in the propagating mode, the optical properties of interference coatings including one or more NIM layers.

\section{The Perfect Lens}

Before we can systematically investigate the reflection and transmission of a plane wave by multilayer thin films containing both positively and negatively refracting materials, we start by analyzing the specific case of the \textit{Perfect Lens} \cite{Pendry_2000} with the help of the admittance formalism detailed in the previous section. 

A perfect lens can be described as a slab (in other words a single layer) of -1 index material with $\epsilon_r'=\mu_r'=-1$ and surrounded by air ($n_s=n_0$, $\epsilon_r'=\mu_r'=1$). At any AOI $\theta_0$, we have 
\begin{equation}
\left\{
\begin{aligned}
&\tilde{n}_{\text{NIM}}^{\text{TE}}=\frac{\cos\theta_0}{\eta_0}=\tilde{n}_0^{\text{TE}}=\tilde{n}_s^{\text{TE}}\\
&\tilde{n}_{\text{NIM}}^{\text{TM}}=\frac{1}{\eta_0\cos\theta_0}=\tilde{n}_0^{\text{TM}}=\tilde{n}_s^{\text{TM}}
\end{aligned}
\right.
\end{equation}
and then following (\ref{eq:RecurrenceRelation})
\begin{equation}
Y_0=\frac{\tilde{n}_0\cos\delta_{\text{NIM}}-i\tilde{n}_0\sin\delta_{\text{NIM}}}{\cos\delta_{\text{NIM}}-i(\tilde{n}_0/\tilde{n}_0)\sin\delta_{\text{NIM}}}=\tilde{n}_0
\end{equation}
regardless of the slab thickness $t_{\text{MIN}}$. This implies that the reflection coefficient $r$ is equal to zero at any angle of incidence and state of polarization (SOP), while the law of conservation of the spatial pulsation takes the form
\begin{multline}
\sigma=\frac{2\pi}{\lambda}n_0\sin\theta_0\\
=\frac{2\pi}{\lambda}n_{\text{NIM}}\sin\theta_{\text{NIM}}=\frac{2\pi}{\lambda}n_s\sin\theta_s
\end{multline}
which imposes : $\theta_{\text{NIM}}=-\theta_0$ and $\theta_s=\theta_0$.

Thanks to the admittance formalism, we also recover very easily two of the three fundamental properties of the \textit{Perfect Lens}, i.e. its antireflection behavior and its imaging ability. Details of the third fundamental property (evanescent wave amplification \cite{Pendry_2000}) are not required to understand the optical properties of multilayer optics in the propagating mode; thus they are not discussed here.

\section{Optical properties of NIM/PIM stacks}
\label{sec:IdealConfigurations}

In this work, we consider ideal negative index materials, i.e. \textbf{without absorption and spectral dispersion}, and above all, without any restriction on the choice of NIM electromagnetic properties.

Moreover, in this section, all the computations are made in the propagating mode, and thus, all the effective indices and phase delays are real.

\subsection{All-NIM configurations}
\label{sec:AllPIM}
Let us consider an all-NIM multilayer stack (refractive index $-n_j$, physical thickness $t_j$, $j=1,2,\dots,p$) deposited on the surface of a PIM semi-infinite substrate (refractive index $n_s$).

The complex admittance $Y_0^{\text{NIM}}$ of this all-NIM stack can be computed with the recursive formula (\ref{eq:RecurrenceRelation}) and the initialization condition (\ref{eq:RecurrenceInitialization})
\begingroup
\small
\begin{equation}
\begin{aligned}
&Y_{p-1}^{\text{NIM}}=\frac{\tilde{n}_s\cos\delta_p^{\text{NIM}}-i\tilde{n}_p\sin\delta_p^{\text{NIM}}}{\cos\delta_p^{\text{NIM}}-i(\tilde{n}_s/\tilde{n}_p)\sin\delta_p^{\text{NIM}}}\\
&Y_{p-2}^{\text{NIM}}=\frac{Y_{p-1}^{\text{NIM}}\cos\delta_{p-1}^{\text{NIM}}-i\tilde{n}_{p-1}\sin\delta_{p-1}^{\text{NIM}}}{\cos\delta_{p-1}^{\text{NIM}}-i(Y_{p-1}^{\text{NIM}}/\tilde{n}_{p-1})\sin\delta_{p-1}^{\text{NIM}}}\\
&\cdots\\
&Y_0^{\text{NIM}}=\frac{Y_1^{\text{NIM}}\cos\delta_1^{\text{NIM}}-i\tilde{n}_1\sin\delta_1^{\text{NIM}}}{\cos\delta_1^{\text{NIM}}-i(Y_1^{\text{NIM}}/\tilde{n}_1)\sin\delta_1^{\text{NIM}}}
\end{aligned}
\end{equation}
\endgroup
As stressed in Section \ref{sec:PIMMirrorImage}, we can replace each NIM layer number by its correlated PIM Mirror Image (refractive index $n_j$, physical thickness $-t_j$, $j=1,2,\dots,p$) and write
\begingroup
\small
\begin{equation}
Y_{p-1}^{\text{NIM}}=\frac{\tilde{n}_s\cos\delta_p+i\tilde{n}_p\sin\delta_p}{\cos\delta_p+i(\tilde{n}_s/\tilde{n}_p)\sin\delta_p}=[Y_{p-1}^{\text{PIM}}]^*
\end{equation}
\begin{equation}
\begin{aligned}
Y_{p-2}^{\text{NIM}}&=\frac{[Y_{p-1}^{\text{PIM}}]^*\cos\delta_{p-1}+i\tilde{n}_{p-1}\sin\delta_{p-1}}{\cos\delta_{p-1}+i([Y_{p-1}^{\text{PIM}}]^*/\tilde{n}_{p-1})\sin\delta_{p-1}}\\
&=[Y_{p-2}^{\text{PIM}}]^*
\end{aligned}
\end{equation}
\begin{equation}
\cdots
\notag
\end{equation}
\begin{equation}
Y_0^{\text{NIM}}=\frac{[Y_1^{\text{PIM}}]^*\cos\delta_1+i\tilde{n}_1\sin\delta_1}{\cos\delta_1+i([Y_1^{\text{PIM}}]^*/\tilde{n}_1)\sin\delta_1}=[Y_{0}^{\text{PIM}}]^*
\end{equation}
\endgroup
where $Y_{j}^{\text{PIM}}$ are the complex admittances of the correlated classical all-PIM multilayer stack (refractive index $n_j$, physical thickness $t_j$).

As a consequence
\begin{equation}
r_{\text{NIM}}=\frac{\tilde{n}_0-[Y_{0}^{\text{PIM}}]^*}{\tilde{n}_0+[Y_{0}^{\text{PIM}}]^*}=r_{\text{PIM}}^*
\end{equation}
which means that the only difference between an all-NIM multilayer configuration and the corresponding all-PIM one is a sign change in the phase of the reflection coefficient. Therefore, the use of all-NIM stacks does not provide any effective benefit with respect to the related all-PIM configurations.

\subsection{Quarter-wavelength Bragg mirror}

\subsubsection{Standard PIM configuration}
Let us consider a standard quarter-wavelength mirror including seven high and low index alternated layers deposited on a semi-infinite glass substrate ($n_s=1.52$) with air as the incident medium ($n_0=1.00$). This stack can be described by the following synthetic formula
\begin{center}
Air / HLHLHLH / Glass
\end{center}
where H (respectively L) is a high-index (respectively low-index) quarter-wavelength layer at the central wavelength of the mirror.

\begin{figure}[htbp]
\centerline{\includegraphics[width=0.8\columnwidth]{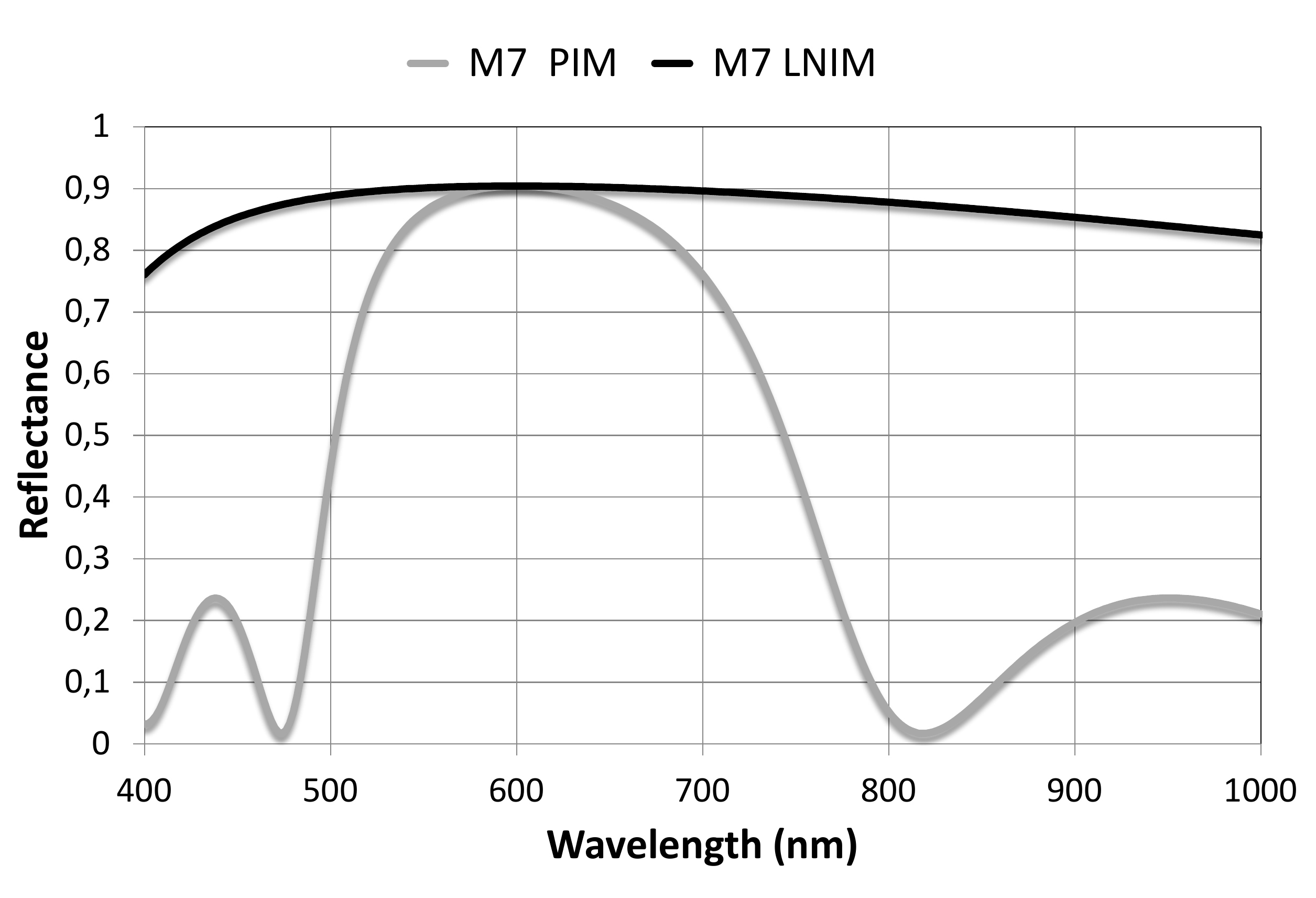}}
\caption{Spectral dependence of the reflectance of seven-layer quarter-wavelength Bragg mirrors (design wavelength = 600 nm). Gray line: M7 PIM; black line: M7 LNIM}
\label{fig:M7Lambda}
\end{figure}

For our simulations, we choose $\lambda_0=600$ nm as the design wavelength, silica ($\text{SiO}_2$, $n_L=1.48$) as the low index material, and tantalum pentoxide ($\text{Ta}_2\text{O}_5$, $n_H=2.24$) as the high index material. The gray line on Fig. \ref{fig:M7Lambda} shows the spectral variation of the reflectance $R=|r|^2$ of this mirror at zero AOI.

\subsubsection{NIM/PIM configurations}
\label{sec:M7NIMPIMPhase}
\paragraph{Low-index NIM layers}
If we replace the low-index L layers of the previous all-PIM quarter-wavelength Bragg mirror by NIM layers ($\bar{L}$, $n_{\bar{L}}$ = -1.48, $t_{\bar{L}}$ =  101.35 nm), the formula of the stack becomes
\begin{center}
Air / H$\bar{L}$H$\bar{L}$H$\bar{L}$H / Glass
\end{center}
and the computation of its optical properties can be achieved by replacing these NIM layers by their correlated PIM Mirror Images ($n_L$ = 1.48, $t_L$ = -101.35 nm). The black line in Fig. \ref{fig:M7Lambda} shows the spectral variation of the reflectance of this M7 LNIM mirror at zero AOI. We observe a spectacular increase in the spectral bandwidth of this mirror induced by the use of low-index NIM layers. This bandwidth now covers almost the entire wavelength range, from 400 to 1000 nm, whereas the spectral width of the corresponding PIM configuration does not exceed 250 nm.
\medskip
\paragraph{High-index NIM layers}

\begin{figure}[htbp]
\centerline{\includegraphics[width=0.8\columnwidth]{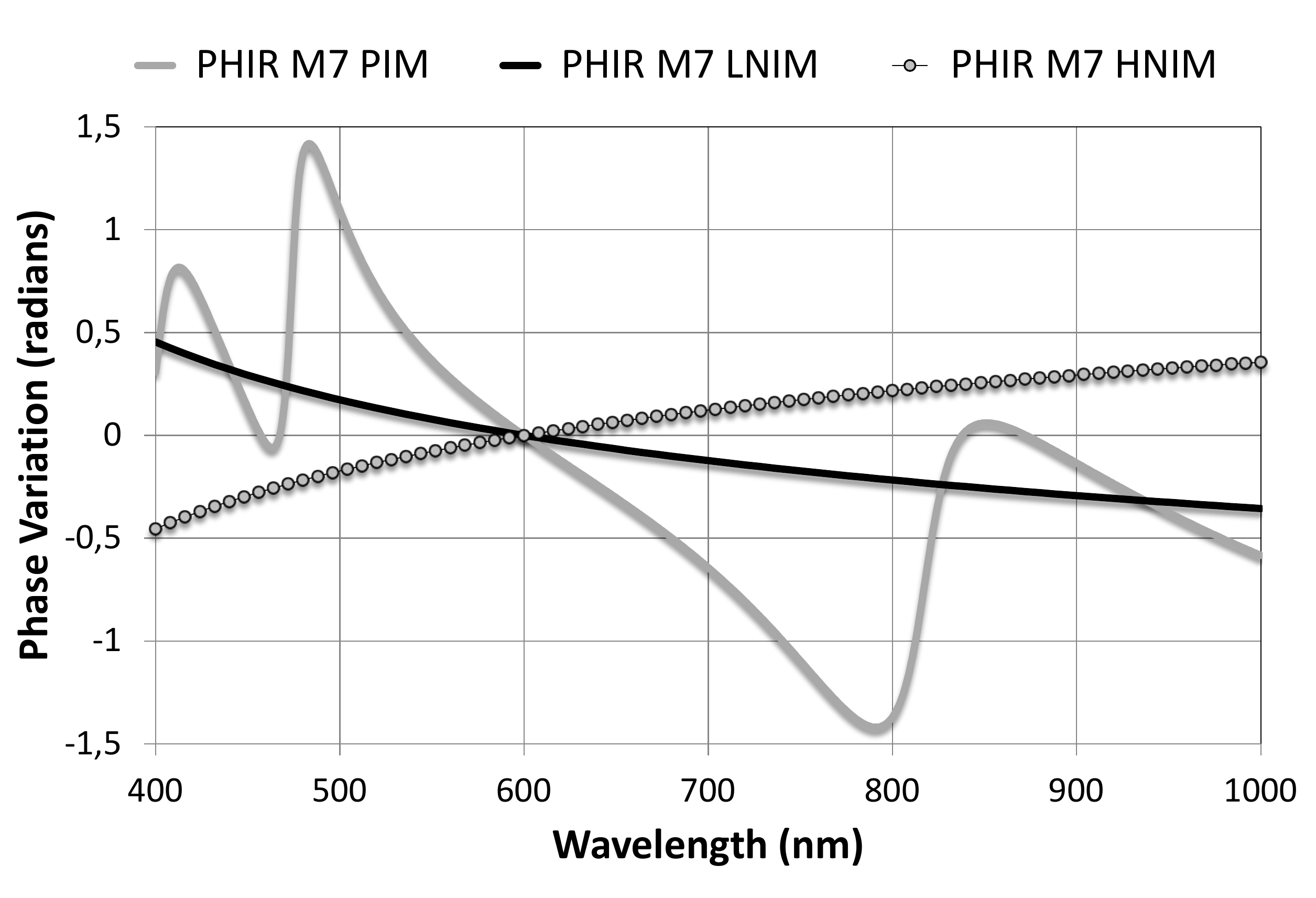}}
\caption{Spectral dependence of the phase at the reflection of seven-layer quarter-wavelength Bragg mirrors (design wavelength = 600 nm). Thick gray line: M7 PIM; thick black line: M7 LNIM; gray circles with black contour: M7 HNIM)}
\label{fig:M7PhiR}
\end{figure}

For a stack formula
\begin{center}
Air / $\bar{H}$L$\bar{H}$L$\bar{H}$L$\bar{H}$ / Glass
\end{center}
where $\bar{H}$ is a high-index quarter-wavelength NIM layer ($n_H$ = -2.24, $t_H$ = 66.96 nm), the result of the simulation (spectral dependence of the reflectance) is exactly the same as that obtained for the low-index NIM layers configuration. The only difference is related to the phase change at the reflection $\varphi$, defined by $r=\sqrt{R}\exp(i\varphi)$ whose spectral dependence is shown in Fig. \ref{fig:M7PhiR} (gray disks with black contour) together with those of the M7 All-PIM (gray line) and M7 LNIM (black line) configurations. The spectral dependence of this phase quantity is indeed \textbf{equal and opposite} to that of the low-index NIM layers configuration (with four high-index NIM layers in this last configuration compared to three low-index NIM layers in the previous one).

\subsection{Fabry-Perot bandpass filter}

\subsubsection{Standard PIM configuration}
\begin{figure}[htbp]
\centerline{\includegraphics[width=0.8\columnwidth]{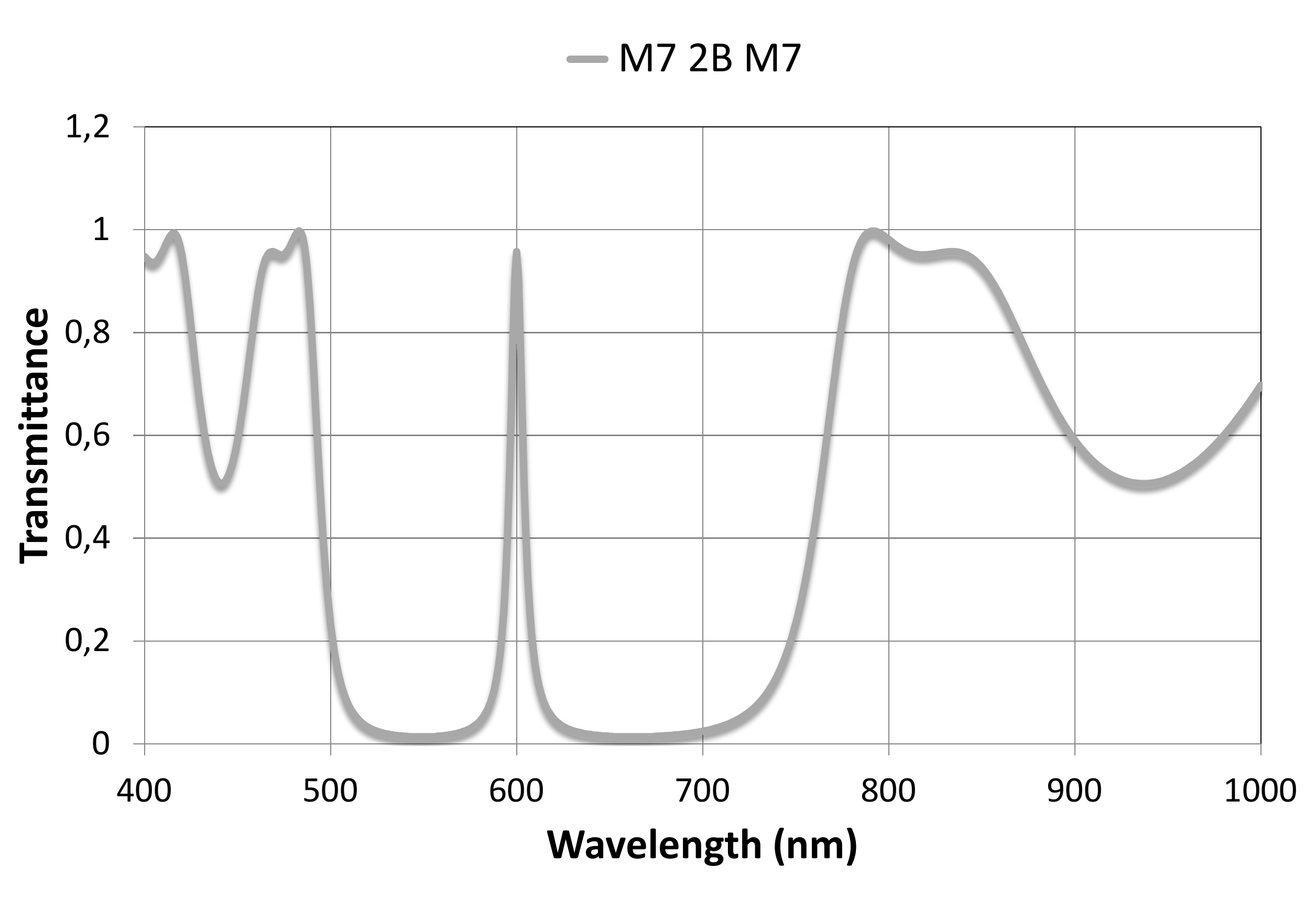}}
\caption{Spectral dependence of the transmittance $T$ of a standard Fabry-Perot bandpass filter (M7 2B M7; all-PIM layers)}
\label{fig:FPPIMLambda}
\end{figure}

Let us now consider a Fabry-Perot bandpass filter including two identical M7 mirrors surrounding a 2B spacer layer. The corresponding multilayer stack is then described by the following formula
\begin{center}
Air / HLHLHLH 2B HLHLHLH / Glass
\end{center}
where H and B have the same meaning as above. The spectral dependence of the transmittance $T$ of this bandpass filter is shown in Fig. \ref{fig:FPPIMLambda}.

The full width at half maximum (FWHM) of the passband $\Delta\lambda$ is about 8 nm whereas that of the stopband is around 150 nm with a maximum rejection level $T_{\text{min}}$ of -18.6 dB.

\subsubsection{NIM/PIM configurations}
\begin{figure}[htbp]
\centerline{\includegraphics[width=0.8\columnwidth]{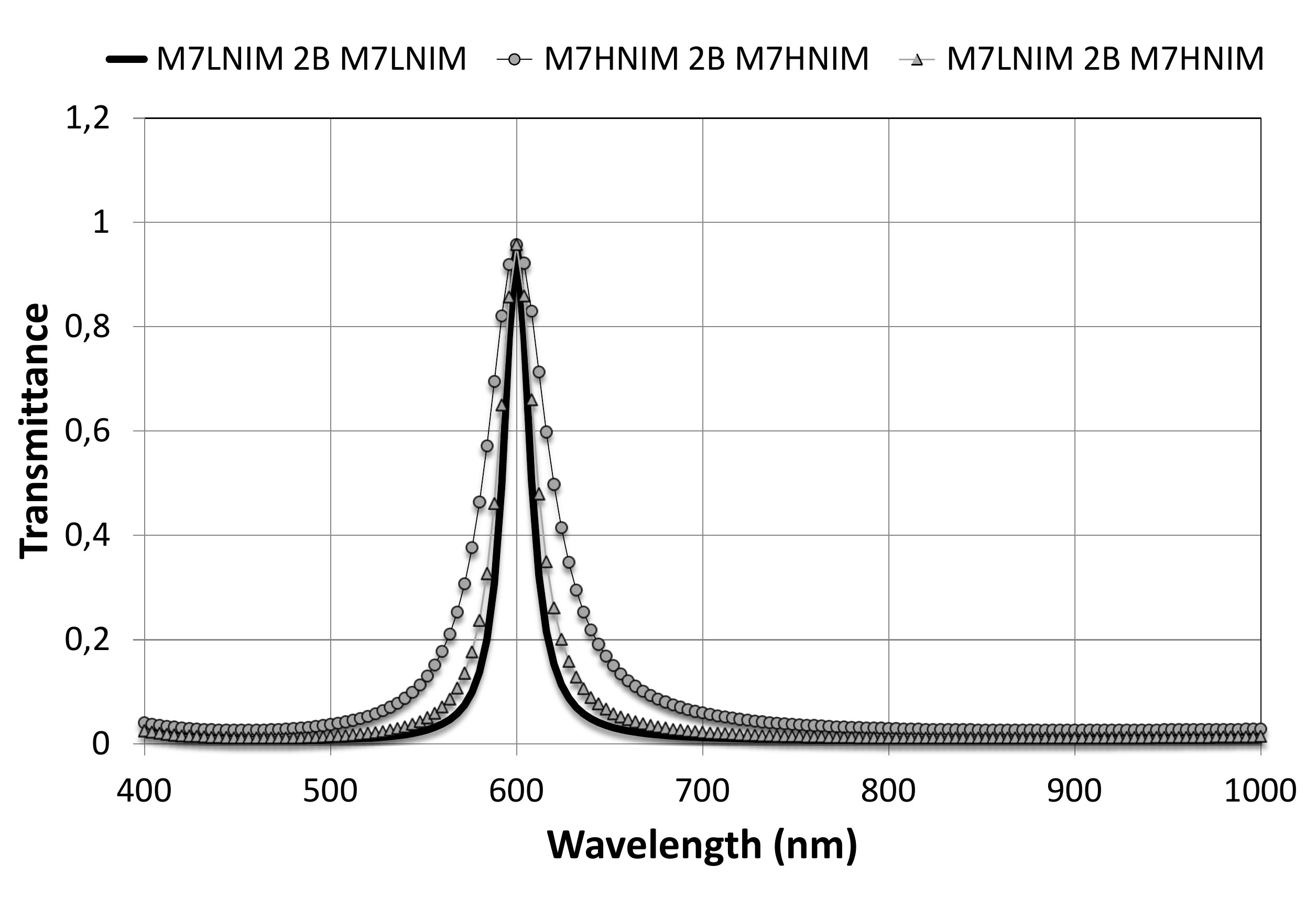}}
\caption{Spectral dependence of the transmittance $T$ of three different Fabry-Perot NIM/PIM configurations. Thick black line: M7LNIM 2B M7LNIM; gray circles with thin black line: M7HNIM 2B M7HNIM; gray triangles with thin gray line: M7LNIM 2B M7HNIM}
\label{fig:FPNIMPIMLambda}
\end{figure}

\paragraph{First analysis}
Taking into account the difference in phase behavior of the NIM and PIM quarter-wavelength Bragg mirrors discussed in Section \ref{sec:M7NIMPIMPhase}, we have to first consider three different NIM/PIM Fabry-Perot configurations, i.e.
\begin{itemize}
\item the M7LNIM 2B M7LNIM configuration, in which the low-index quarter-wavelength PIM layers of both M7 mirrors are replaced by corresponding NIM layers,
\item the M7HNIM 2B M7HNIM configuration, in which the NIM layers are now the high-index layers of both M7 mirrors, and
\item the M7LNIM 2B M7HNIM configuration, which corresponds to a design intermediate to the two previous ones.
\end{itemize}
The spectral dependence of the transmittance of these three Fabry-Perot configurations is shown in Fig. \ref{fig:FPNIMPIMLambda}.

The replacement of some PIM layers by NIM layers in the standard design of this Fabry-Perot (FP) bandpass filter induces then, for the three considered configurations, a large increase in the width of the stopband (more than a factor of 4) which allows us to avoid the use of blocking filters for most potential applications. However the spectral properties of these three configurations are not exactly identical, and, in order to highlight their differences, we have summarized in Table \ref{tab:ComparisonFP}, for each FP configuration, the values of the two parameters introduced in the previous section, i.e. the filter bandwidth $\Delta\lambda$ (FWHM) and the maximum rejection level $T_{\text{min}}$ in the stopband.

\begin{table}[htbp]
\caption{\label{tab:ComparisonFP}Comparison of FP spectral properties.}
\begin{ruledtabular}
\begin{tabular}{lcc}
FP stack formula & $\Delta\lambda$ (nm) & $T_{\text{min}}$ (dB)\\
$(\text{HL})^3$H 2B H$(\text{LH})^3$ & 8.5 & -18.6\\
$(\text{H$\bar{L}$})^3$H 2B H$(\text{$\bar{L}$H})^3$ & 16.8 & -21.3\\
$(\text{$\bar{H}$L})^3$$\bar{H}$ 2B $\bar{H}$$(\text{L$\bar{H}$})^3$ & 40.3 & -15.8\\
$(\text{H$\bar{L}$})^3$H 2B $\bar{H}$$(\text{L$\bar{H}$})^3$ & 23.6 & -19.5\\
\end{tabular}
\end{ruledtabular}
\end{table}

The bandwidth of these Fabry-Perot filters is between 8.5 nm (for the standard configuration) and 40.3 nm (for the configuration using high-index quarter-wavelength NIM layers), whereas the rejection level is not drastically affected by these design changes (being between -15.8 dB and -21.3 dB).
\medskip
\paragraph{Discussion}
To explain the differences in behavior observed among these various configurations, we must consider the overall round-trip phase $\Phi$ of each Fabry-Perot configuration and study its spectral dependence. We have indeed at normal incidence
\begin{equation}
\Phi(\lambda)=2\pi\frac{\Delta}{\lambda}+\rho_{\text{US}}^-(\lambda)+\rho_{\text{LS}}^+(\lambda)
\end{equation}
where $\Delta$ is the optical path difference corresponding to the round-trip of light in the spacer layer and $\rho_{\text{US}}^-$ (respectively $\rho_{\text{LS}}^+$) is the phase change at the reflection of light on the upper mirror (respectively lower mirror) of the Fabry-Perot cavity. This means that, for an M7 2B M7 configuration, we have to consider for the $\rho_{\text{LS}}^+$ term a seven-layer quarter-wavelength Bragg mirror deposited on a glass substrate with \textbf{silica as incident medium} and for the  $\rho_{\text{US}}^-$ term a seven layers quarter-wavelength Bragg mirror deposited \textbf{on air} with silica, again, as the incident medium.

The quantity $\Delta$ is given here by 
\begin{equation}
\Delta=2n_{\text{sp}}e_{\text{sp}}
\end{equation}
where the index sp is related to the \textit{spacer} of the Fabry-Perot cavity. At the design wavelength, we thus have \cite{Macleod_2010}
\begin{equation}
\Phi(\lambda_0)=\frac{4\pi n_{\text{sp}}e_{\text{sp}}}{\lambda_0}+\rho_{\text{US}}^-(\lambda_0)+\rho_{\text{LS}}^+(\lambda_0)=2p\pi
\end{equation}
where $p$ is in $\mathbb{Z}$.

In general, the transmittance $T(\lambda)$ of a Fabry-Perot multilayer filter around the design wavelength $\lambda_0$ is given by
\begin{equation}
T(\lambda)=\frac{T(\lambda_0)}{1+A\sin^2\frac{\Phi(\lambda)}{2}}
\end{equation}
with
\begin{equation}
A=\frac{4\sqrt{R_{\text{LS}}^+R_{\text{US}}^-}}{(1-\sqrt{R_{\text{LS}}^+R_{\text{US}}^-})^2}
\end{equation}
where $R_{\text{US}}^-$ (respectively $R_{\text{LS}}^+$) is the reflectance of the upper mirror (respectively lower mirror) of the Fabry-Perot cavity. Accordingly, the bandpass width $\Delta\lambda$ is defined by
\begin{equation}
\Phi(\lambda_0\pm\frac{\Delta\lambda}{2})=\frac{1-\sqrt{R_{\text{LS}}^+R_{\text{US}}^-}}{\sqrt[4]{R_{\text{LS}}^+R_{\text{US}}^-}}
\end{equation}
This means that the bandwidth of the Fabry-Perot is directly driven by the \textbf{spectral dependence of the overall round-trip phase} $\Phi$. Moreover, in the linear regime, we can write \cite{Lequime_2009}
\begin{equation}
\Delta\lambda\approx\frac{2}{|\left[\frac{\partial\Phi}{\partial\lambda}\right]_{\lambda_0}|}\cdot\frac{1-\sqrt{R_{\text{LS}}^+R_{\text{US}}^-}}{\sqrt[4]{R_{\text{LS}}^+R_{\text{US}}^-}}
\end{equation}
where the linear dependence of the overall round-trip phase around the design wavelength $\lambda_0$ is given by
\begin{equation}
\left[\frac{\partial\Phi}{\partial\lambda}\right]_{\lambda_0}=-\frac{4\pi n_{\text{sp}}e_{\text{sp}}}{\lambda_0^2}+\left.\frac{\partial\rho_{\text{US}}^-}{\partial\lambda}\right|_{\lambda_0}+\left.\frac{\partial\rho_{\text{LS}}^+}{\partial\lambda}\right|_{\lambda_0}
\label{eq:PhiLinearDependence}
\end{equation}
For all the Fabry-Perot configurations considered in Table \ref{tab:ComparisonFP}, the spacer layer is half-wave at the design wavelength, whereby relation (\ref{eq:PhiLinearDependence}) becomes
\begin{equation}
\left[\frac{\partial\Phi}{\partial\lambda}\right]_{\lambda_0}=-\frac{2\pi}{\lambda_0}+\left.\frac{\partial\rho_{\text{US}}^-}{\partial\lambda}\right|_{\lambda_0}+\left.\frac{\partial\rho_{\text{LS}}^+}{\partial\lambda}\right|_{\lambda_0}
\end{equation}
Thus, minimizing the linear dependence of the overall round-trip phase $\Phi$ (and accordingly maximizing the related spectral bandwidth $\Delta\lambda$) requires the use of mirrors characterized by a \textbf{positive} spectral dependence of the global phase change at the reflection for the design wavelength.
\begin{table}[!h]
\caption{\label{tab:SpectralDispersionData}FP spectral dispersion properties (in $\text{nm}^{-1}$) and corresponding bandwidth predicted by the linear approach (in nm).}
\begin{ruledtabular}
\begin{tabular}{lcccc}
FP stack formula & $\left.\frac{\partial\rho_{\text{LS}}^+}{\partial\lambda}\right|_{\lambda_0}$ & $\left.\frac{\partial\rho_{\text{US}}^-}{\partial\lambda}\right|_{\lambda_0}$ & $\left[\frac{\partial\Phi}{\partial\lambda}\right]_{\lambda_0}$ & $\Delta\lambda$\\
$(\text{HL})^3$H 2B H$(\text{LH})^3$ & -0.00929 & -0.00950 & -0.02926 & 8.5\\
$(\text{H$\bar{L}$})^3$H 2B H$(\text{$\bar{L}$H})^3$ & -0.00216 & -0.00212 & -0.01475 & 16.8\\
$(\text{$\bar{H}$L})^3$$\bar{H}$ 2B $\bar{H}$$(\text{L$\bar{H}$})^3$ & 0.00216 & 0.00212 & -0.00619 & 40.1\\
$(\text{H$\bar{L}$})^3$H 2B $\bar{H}$$(\text{L$\bar{H}$})^3$ & -0.00216 & 0.00212 & -0.01052 & 23.6\\
\end{tabular}
\end{ruledtabular}
\end{table}

The spectral bandwidths predicted by using this approach are summarized in Table \ref{tab:SpectralDispersionData} for the four previous  Fabry-Perot configurations. They are in perfect accordance with those given in Table \ref{tab:ComparisonFP}.

The same approach (minimizing the spectral dependence of the overall round trip phase $\Phi$) can be used to design a kind of \textit{white} Fabry-Perot cavity, i.e. a multilayer cavity that spontaneously exhibits a resonant behavior over a very large spectral range. The description of this optimization scheme exceeds the frame of this paper and will be described further in a dedicated article \cite{Lequime}.

\subsection{Antireflection coatings}

\subsubsection{Standard PIM configurations}

We restrict our study to single-layer and double-layer antireflection stacks providing zero reflectance at a given design wavelength (here 600 nm) and normal incidence. Theoretical analyses using admittance formalism \cite{Macleod_2010} shows that we have only two possible configurations 
\begin{itemize}
\item a single-layer solution (1A) whose refractive index and optical thickness are imposed ($n_1=\sqrt{n_0n_s}$, $n_1t_1=\lambda/4$ at the design wavelength), or
\item a double-layer solution for which the phase delays of both layers must satisfy the following set of equations
\begin{equation}
\left\{
\begin{aligned}
&\tan^2\delta_1=\frac{\tilde{n}_1^2(\tilde{n}_s-\tilde{n}_0)(\tilde{n}_2^2-\tilde{n}_0\tilde{n}_s)}{(\tilde{n}_0\tilde{n}_2^2-\tilde{n}_s\tilde{n}_1^2)(\tilde{n}_1^2-\tilde{n}_0\tilde{n}_s)}\\
&\tan\delta_2=\frac{\tilde{n}_1\tilde{n}_2(\tilde{n}_0-\tilde{n}_s)}{(\tilde{n}_0\tilde{n}_2^2-\tilde{n}_s\tilde{n}_1^2)}\cdot\frac{1}{\tan\delta_1}
\end{aligned}
\right.
\label{eq:TwoLayerAR}
\end{equation}
In this case, we can either choose the same low-index and high-index materials as above, i.e. $\text{SiO}_2$ ($n_1=1.48$) and $\text{Ta}_2\text{O}_5$ ($n_2=2.24$) and thereby define two different solutions following the sign of $\tan\delta_1$

\smallskip
Solution 2A: Air / 1.312L 0.317H / Glass

Solution 2B: Air / 0.688L 1.683H / Glass

\smallskip
or we can impose that the optical thicknesses of both layers be equal to $\lambda/4$, which leads to the following relation
\begin{equation}
n_0n_2^2-n_sn_1^2 = 0
\end{equation}
If we use silica for the top layer ($n_1=1.48$), the refractive index $n_2$ of the bottom layer will be equal to 1.825, the formula of the corresponding stack being simply given in this case by

\smallskip
Solution 2C: Air / L H / Glass
\end{itemize}

\begin{figure}[htbp]
\centerline{\includegraphics[width=0.8\columnwidth]{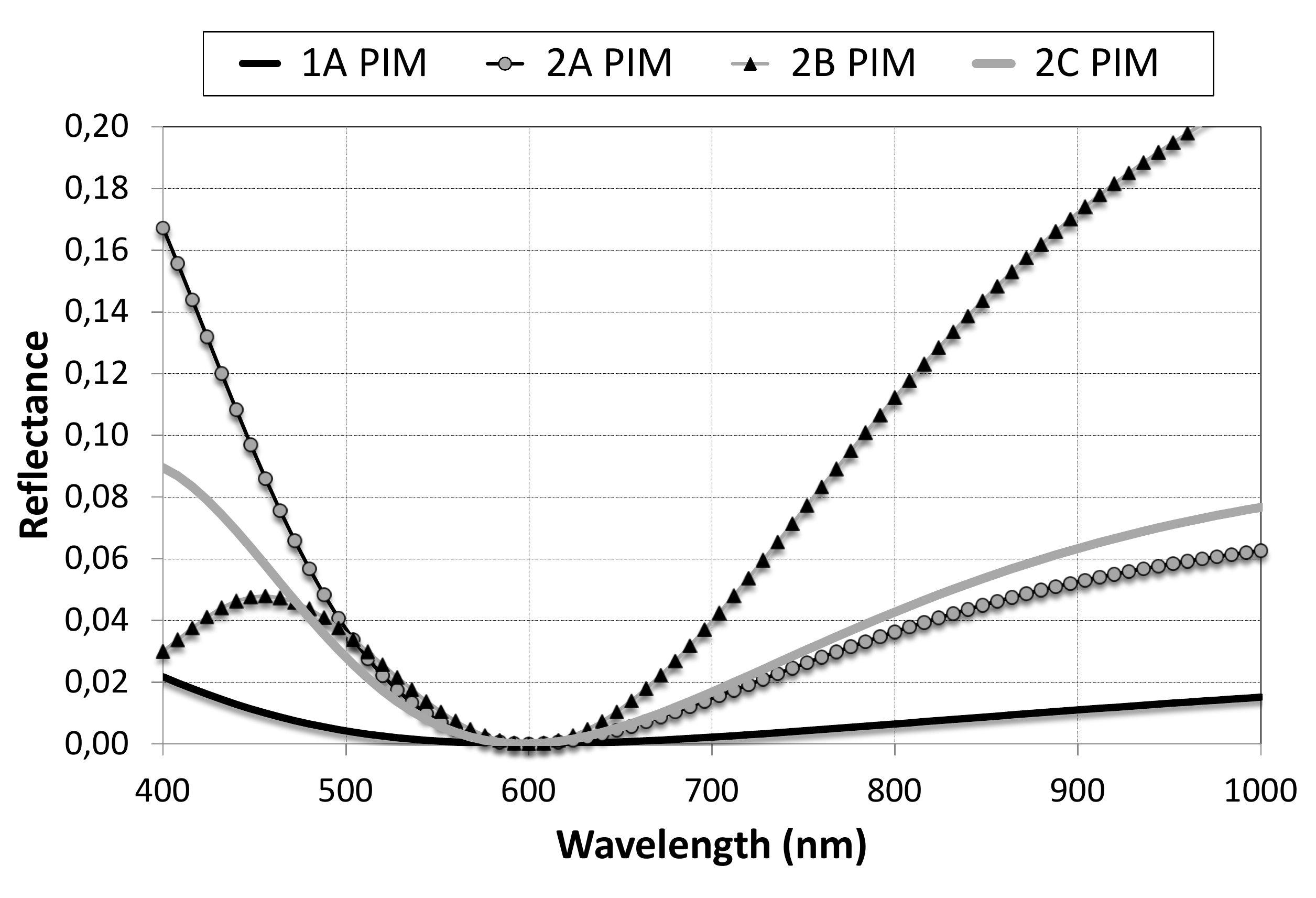}}
\caption{Spectral dependence of the reflectance of all-PIM single-layer and double-layer antireflection coatings optimized at 600 nm. Thick black line: solution 1A; gray disks with thin black line: solution 2A; black triangles with thin gray line: solution 2B; thick gray line: solution 2C.}
\label{fig:ARPIMLambda}
\end{figure}

\smallskip
The Figure \ref{fig:ARPIMLambda} shows the spectral dependence of the reflectance $R=|r|^2$ of the single-layer solution (1A) as well as the three double-layer solutions (2A, 2B and 2C) at zero AOI. Clearly the single-layer solution is the best one in terms of behavior throughout the spectral range.

\subsubsection{NIM/PIM configurations}
In accordance with the conclusion of Section \ref{sec:AllPIM}, we do not need to analyze the optical properties of all-NIM antireflection coatings, because the spectral dependence of their reflectance $R$ would be identical to those shown in Fig. \ref{fig:ARPIMLambda}.

Hence we focus our further analysis on the NIM/PIM double-layer configurations by considering only the stacks in which the high-index PIM layer is replaced by a NIM layer (and, except for the phase change at reflection, the replacement of the only low-index layer would provide obviously identical results). In this case, the set of equations  (\ref{eq:TwoLayerAR}) remains exactly the same, but the results in terms of physical thickness $t_2$ must now be \textbf{negative}. The three NIM/PIM antireflection double-layer stacks are thus described by the formulas

\smallskip
Solution 2$\bar{A}$: Air / 1.312L 1.683$\bar{H}$ / Glass

\smallskip
Solution 2$\bar{B}$: Air / 0.688L 0.317$\bar{H}$ / Glass

\smallskip
Solution 2$\bar{C}$: Air / L $\bar{H}$ / Glass

\begin{figure}[htbp]
\centerline{\includegraphics[width=0.8\columnwidth]{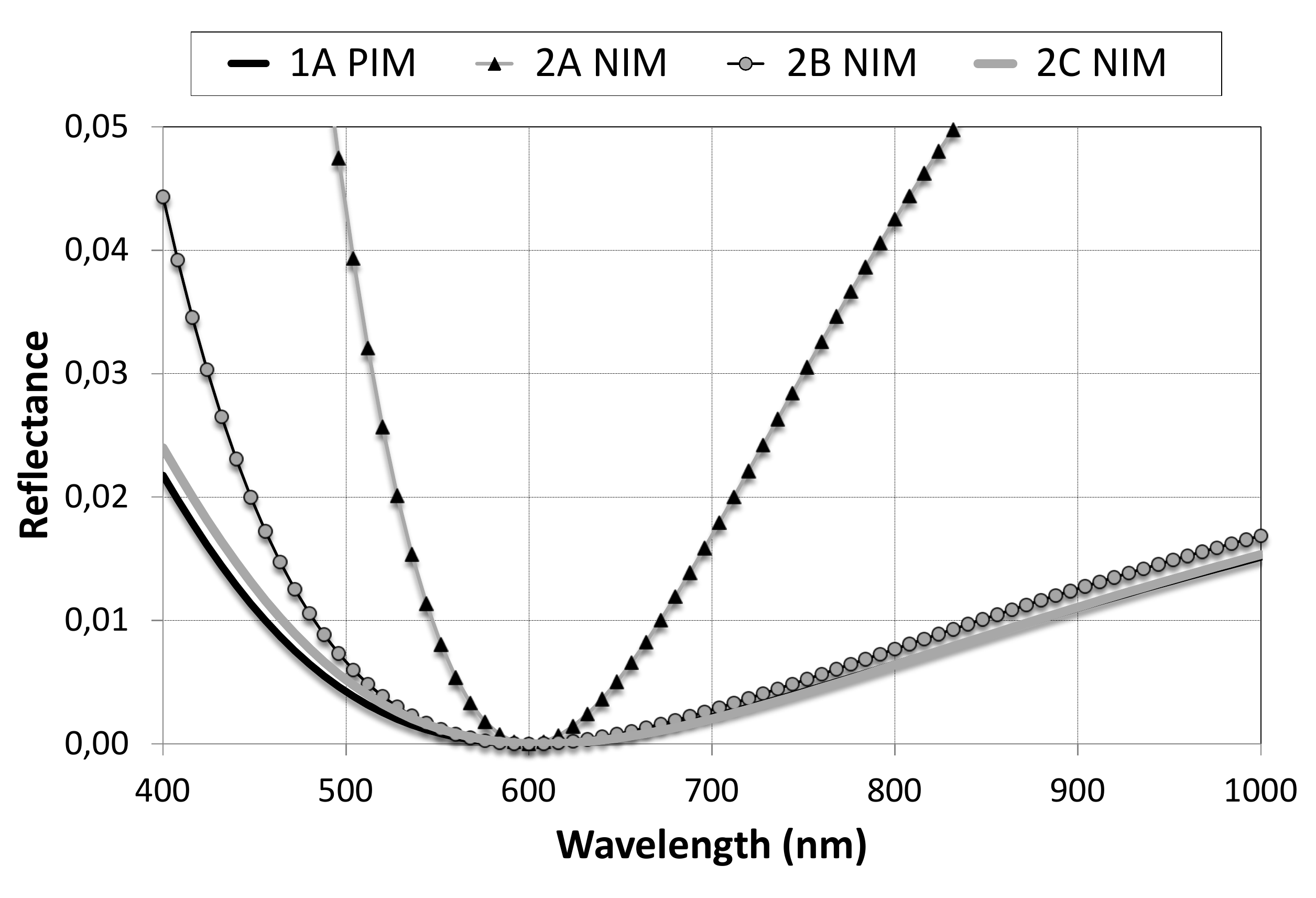}}
\caption{Spectral dependence of the reflectance of PIM/NIM double-layer antireflection coatings optimized at 600 nm. Thick thick line: solution 1A; black triangles with thin gray line: solution 2$\bar{A}$; gray disks with thin black line: solution 2$\bar{B}$; thick gray line: solution 2$\bar{C}$.}
\label{fig:ARNIMLambda}
\end{figure}

\smallskip
and the spectral dependence of their reflectance $R$ is shown in Fig. \ref{fig:ARNIMLambda} (where, in order to exemplify the differences among the various curves, the vertical scale is expanded by a factor of 4 with respect to Fig. \ref{fig:ARPIMLambda}).

We observe that the spectral dependence of the residual reflectance provided by solution 2$\bar{C}$ is nearly identical to that of solution 1A. The use of a NIM layer in a double-layer configuration is then a way to synthesize a virtual index in accordance with the single layer condition. Nevertheless, it does not allow us to achieve \textit{perfect} spectral behavior, i.e., zero reflectance at any wavelength over a wide spectral range (400-1000 nm for instance).

\section{Conclusion}
In this paper, we hope to have conclusively shown the usefulness of the admittance formalism to model the optical properties of multilayer stacks including negative index materials. The most attractive results concern the quarter-wavelength Bragg mirrors and the multilayer Fabry-Perot filters, and the ability to tailor the phase properties of such multilayer structures open interesting avenues for design of \textit{white} Fabry-Perot. Other potential applications of the admittance formalism include the analysis of zero-index stop band structures which consist in an alternation of layers of opposite refractive indices \cite{Li_2003}. Finally, it could be really attractive to apply this approach to the study of optical properties of the same multilayer structures with more realistic spectral material properties, as defined by V. A. Podolskiy \cite{Podolskiy_2003}, V. M. Shalaev \cite{Shalaev_2005} or more recently Z. H. Jiang \cite{Jiang_2013}.

\end{document}